# High Pressure Study of Structural Phase Transitions and Superconductivity in La$_{1.48}$Nd$_{0.4}$Sr$_{0.12}$CuO$_4$

M.K. Crawford[1], R.L. Harlow[1], S. Deemyad[2], V. Tissen[2]*, J.S. Schilling[2], E.M. McCarron[1], S.W. Tozer[3], D.E. Cox[4], N. Ichikawa[5], S. Uchida[5], Q. Huang[6]

[1]*DuPont Co., Central Research and Development, Wilmington, DE 19880-0356*

[2]*Department of Physics, Washington University, St. Louis, MO 63130-4899*

[3]*National High Magnetic Field Laboratory, Tallahassee, FL 32310*

[4]*Department of Physics, Brookhaven National Laboratory, Upton, N.Y. 11973*

[5]*Department of Physics, University of Tokyo, Tokyo 113-8656, Japan*

[6]*NIST Center for Neutron Research, National Institute of Standards and Technology, Gaithersburg, MD 20899*

*Permanent address: Institute of Solid State Physics, Chernogolovka, Russia

We have determined the crystal structures and superconducting transition temperatures of La$_{1.48}$Nd$_{0.4}$Sr$_{0.12}$CuO$_4$ under nearly hydrostatic pressures in diamond anvil cells to 5.0 GPa and 19.0 GPa, respectively. Synchrotron x-ray powder diffraction measurements were used to establish the pressure-temperature structural phase diagram. Under pressure the superconducting transition temperature increases rapidly from T$_c$ ≈ 3 K to a maximum value of 22 K at 5 GPa, a pressure slightly greater than that required to stabilize the undistorted *I4/mmm* structure in the superconducting state. Increasing the pressure further to 19 GPa leads to a decrease in $T_c$ to ~12 K. These results are discussed in relation to earlier high pressure measurements for similar materials.

1. INTRODUCTION

Despite their structural simplicity, there are features unique to the $La_2CuO_4$-based superconductors that provide additional insight concerning the microscopic interactions that lead to high temperature superconductivity. Several different structural phase transitions have been observed in $La_2CuO_4$ doped with alkaline-earth [1] or alkaline-earth and rare-earth metals [2]. As illustrated in Figure 1, these phase transitions are characterized by degenerate order parameters $Q_1$ and $Q_2$ which describe tilting of the copper-oxygen octahedra about the (110) or (1-10) axes of the high-temperature tetragonal (HTT) structure; the four observed tilt structures and the corresponding values for the order parameters are listed in Table I. These structural transformations have a profound effect upon superconductivity [1-4]. For example, the appearance in $La_{1.875}Ba_{0.125}CuO_4$ of the low-temperature tetragonal (LTT) phase, in addition to the previously known low-temperature orthorhombic (LTO1) phase, is associated with nearly complete suppression of superconductivity in that material [1].

Unlike the $La_{2-x}Ba_xCuO_4$ system, however, in materials of composition $La_{2-x-y}Nd_ySr_xCuO_4$ the doping or crystal structure can be independently controlled with changes of $x$ or $y$, respectively [2]. These materials transform directly to the LTT phase, or to a second orthorhombic LTO2 phase (Table I), over a broad range of $x$ and $y$ values. Neutron scattering studies of single crystals of $La_{1.48}Nd_{0.4}Sr_{0.12}CuO_4$ first revealed the presence of static charge and spin stripes in the copper-oxygen planes of a cuprate [5]. The spin stripes are commensurate with the lattice at $x = 1/8$, and it was suggested [5] that the structural distortion present in the LTT phase pins the charge stripes and thus suppresses superconductivity. More recently, charge order coexisting with superconductivity has also been observed in several other underdoped cuprates using scanning tunneling microscopy [6, 7], and a number of theoretical explanations have already been proposed for the these observations [8-10]. To understand the effect of structural distortions on charge ordering and superconductivity in cuprates is an important goal, and in this paper we describe the relationship between structural distortions and superconductivity for $La_{1.48}Nd_{0.4}Sr_{0.12}CuO_4$, a cuprate in which charge order has been observed [5].

In addition to temperature and chemical composition, pressure can be used to control the structural phase transitions and vary $T_c$ in lanthanum cuprates [10]. In this paper we report the effect of nearly hydrostatic pressure on the structure and superconducting transition temperature $T_c$ of $La_{1.48}Nd_{0.4}Sr_{0.12}CuO_4$. In contrast to several earlier studies of the structural [12] and superconducting [13] properties of the related material $La_{2-x}Ba_xCuO_4$, our x-ray diffraction and AC susceptibility measurements of $La_{1.48}Nd_{0.4}Sr_{0.12}CuO_4$ extend to pressures sufficiently high to completely suppress the structural transitions. Our data are thus, to our knowledge, the first to reveal the complete correlation between the structure and $T_c$ in this material, for which the presence of spin and charge order at ambient pressure is fully established [5]. There have also been several recent studies of the effects of hydrostatic and uniaxial pressure on superconductivity in $La_{1.48}Nd_{0.4}Sr_{0.12}CuO_4$ [14] and $La_{1.64}Eu_{0.2}Sr_{0.16}CuO_4$ [15] or $La_{1.65}Eu_{0.2}Sr_{0.15}CuO_4$ [16] (the latter materials also have the LTT structure at low temperatures). These studies did not, however, include direct determinations of the structures of these materials under pressure.

2. EXPERIMENT

Powder samples for x-ray diffraction measurements were synthesized using standard solid-state chemical techniques. The starting materials $La_2O_3$, $Nd_2O_3$, CuO, and SrO were mixed in the appropriate ratios and fired in flowing oxygen at a temperature of 1100 ºC for 12 hours. The resulting sample of $La_{1.48}Nd_{0.4}Sr_{0.12}CuO_4$ was single-phase by x-ray and neutron powder diffraction. Refinement of neutron diffraction data yielded lattice parameters at T = 300 K of $a$ = 5.33219(7) Å, $b$ = 5.36245(6) Å, and $c$ = 13.15202(17) Å for a unit cell of *Bmab* space-group symmetry. A SQUID magnetometer was used to determine the temperature of the superconducting transition at $T_c$ = 4.3 K (midpoint) in a 1 G field (Figure 2). The shielding signal is $\chi$ = –0.07 emu/cm$^3$ for the 38-45 μm particles, whereas for a perfect superconductor one expects $\chi$ = -1/(4π) = -0.08 emu/cm$^3$, ignoring the effect of demagnetization [17]. Thus the polycrystalline sample appears to be nearly 100% superconducting.

High-pressure x-ray powder diffraction data were obtained using a Merrill-Bassett diamond anvil cell. The pressure medium was a 4:1 methanol:ethanol mixture which provides a hydrostatic environment to pressures as high as 10 GPa at room temperature (RT). Pressures were measured at RT using the pressure-induced shift of the R1 fluorescence line from small ruby chips located inside the pressure cell; at low temperatures the known equations of state of NaCl or $CaF_2$, either of which were included in the pressure cell, were used to determine the pressure. The x-ray powder diffraction data were collected at beamline X-7A at the National Synchrotron Light Source at Brookhaven National Laboratory using a position-sensitive detector [18].

Single crystals of $La_{1.48}Nd_{0.4}Sr_{0.12}CuO_4$ were grown using the traveling solvent floating zone technique, and the desired composition confirmed by electron-probe microscope analysis. Two single crystal samples, each having dimensions of approximately 120 x 100 x 30 $\mu m^3$, were removed from a mother crystal of $La_{1.48}Nd_{0.4}Sr_{0.12}CuO_4$ and used in the $T_c(P)$ measurements. The magnetic susceptibility of the randomly oriented mother crystal was measured in a SQUID magnetometer at 1 G field, and the data is shown in Figure 2. The shielding signal *exceeds* the value of $\chi = -1/(4\pi)$ expected for a perfect superconductor, most likely due to the effect of demagnetization; for a superconducting sphere $\chi = -3/(8\pi) = -0.12$ [17], slightly larger than the value we observe for our (nonspherical) single crystal. Thus we believe the single crystal is also fully superconducting. The superconducting transition temperature of the crystal at ambient pressure, $T_c \approx 3.5$ K (midpoint), is slightly lower than that of the polycrystalline sample used in the structural studies.

The structural phase diagram determined for single crystal $La_{1.6-x}Nd_{0.4}Sr_xCuO_4$ [5] is in good agreement with that found earlier by powder x-ray diffraction [2]; thus we are confident that our structural studies of *powders* at high pressure are directly comparable to the superconducting $T_c(P)$ measurements we have made using *single crystals*. The latter were chosen for the $T_c(P)$ determinations, rather than powders, due to their larger AC susceptibility (shielding) signals and sharper superconducting transitions (Figure 2).

The superconducting transition temperatures of the $La_{1.48}Nd_{0.4}Sr_{0.12}CuO_4$ crystals under nearly hydrostatic pressure were determined inductively to ± 0.1 K using an AC susceptibility technique. The high-pressure apparatus utilizes a diamond-anvil cell with $^4$He as pressure medium. The pressure in the cell was determined at any temperature at or below RT to within 0.2 GPa using a ruby manometer. A small piece of Pb was also included in the pressure cell for use as a superconducting manometer [19]. Further details of these high-pressure techniques are given elsewhere [20].

3. RESULTS AND DISCUSSION

In Figure 3 we show an x-ray powder diffraction pattern for $La_{1.48}Nd_{0.4}Sr_{0.12}CuO_4$ measured at a temperature of 10 K and at a pressure of 4.2 GPa. The narrow widths of the Bragg reflections demonstrate that the pressure medium is nearly hydrostatic at the highest pressures and lowest temperatures studied. The x-ray diffraction pattern can be fit [21] assuming that $La_{1.48}Nd_{0.4}Sr_{0.12}CuO_4$ has the HTT structure. This conclusion is primarily based upon the observation that the ($hk0$) reflections are not split, as they are in the LTO1 or LTO2 structures. Furthermore, as shown in Figure 4, the orthorhombic distortion at T = 100 K in the LTO1 phase decreases with pressure at a rate that suggests that the tilt of the $CuO_6$ octahedra will vanish at a pressure somewhat above 4.0 GPa. This observation rules out the possibility that the low temperature structure is LTT rather than HTT at 4.2 GPa.

In Figure 3 we also show the diffraction pattern for $La_{1.48}Nd_{0.4}Sr_{0.12}CuO_4$ measured at a pressure of 2.2 GPa. In this case the structure is orthorhombic at 10 K. These data allow a choice to be made between the LTO1 or LTO2 structure. In Figure 5 we show the $(110)_{HTT}$ Bragg reflection measured at a pressure of 2.2 GPa at three temperatures. At 300 K the structure is HTT (not shown), but at 100 K the $(110)_{HTT}$ peak splits into the $(200)_{LTO1}$ and $(020)_{LTO1}$ peaks characteristic of the LTO1 structure. After cooling to 75 K, two central peaks appear between the LTO1 peaks; these we assign to the LTO2 phase. The LTO2 phase coexists with the LTO1 phase over a small temperature range, demonstrating that the structural transition is first-order, as it is at ambient pressure [2]. At 10 K only the two central peaks characteristic of the LTO2 structure remain.

In Figure 4 we plot the orthorhombic strain versus temperature for $La_{1.48}Nd_{0.4}Sr_{0.12}CuO_4$ at five different pressures. The HTT → LTO1 structural transition is gradually suppressed by pressure, and vanishes at pressures somewhat greater than 4.0 GPa. We also show in this figure the $CuO_6$ octahedra tilt angles as a function of pressure, estimated from the orthorhombic strains at T = 100 K. Our data can also be used to determine the values of the linear compressibilities of the HTT phase of $La_{1.48}Nd_{0.4}Sr_{0.12}CuO_4$ at T = 300 K: $\kappa_a = (2.85 \pm 0.31) \times 10^{-3}$ $GPa^{-1}$ and $\kappa_c = (2.07 \pm 0.33) \times 10^{-3}$ $GPa^{-1}$. These values are similar to those reported for other lanthanum cuprates [11].

The results of our high-pressure x-ray diffraction experiments are summarized in Figure 6 (top), which presents the (T, P) structural phase diagram of $La_{1.48}Nd_{0.4}Sr_{0.12}CuO_4$. The phase diagram exhibits several structural phases (LTO1, LTO2, and LTT) as a function of temperature at ambient and low pressures, but with increasing pressure first the LTT phase is eliminated, and above 4.0 GPa only the HTT phase is observed. In the $La_{1.48}Nd_{0.4}Sr_{0.12}CuO_4$ system it is known [2-4] that the LTT and the LTO2 phases strongly suppress superconductivity, leading to anomalously low ambient pressure $T_c$ values. According to the structural phase diagram in Figure 6 (top), as the applied pressure is increased from ambient to above 4.0 GPa, the LTT and LTO2 phases vanish and the $T_c$ should increase by as much as an order of magnitude.

In Figure 6 (bottom) we show the results of two consecutive high-pressure measurement series on $La_{1.48}Nd_{0.4}Sr_{0.12}CuO_4$ to pressures as high as 19 GPa. To experimental accuracy, $T_c(P)$ is identical for both experiments and is seen to be reversible in pressure over the entire pressure range. For the second experiment (unprimed numbers) it is interesting to note that, as the pressure is released below 6 GPa, the superconducting transition broadens markedly. This may signal that the sample has entered a multiphase region as the HTT phase converts to the LTO2 and LTT phases.

As the pressure is increased to 5.0 GPa, $T_c$ is seen to increase at the rapid rate of approximately +4.1 K/GPa, presumably due to the suppression of the LTT and LTO2 phases with pressure. In $La_{1.875}Ba_{0.125}CuO_4$, $dT_c/dP = +10(\pm 2)$ K/GPa [11-13], more than twice the value we measure in $La_{1.48}Nd_{0.4}Sr_{0.12}CuO_4$. We expect that $dT_c/dP$ will scale *inversely* with the ambient pressure $CuO_6$ tilt angle, and this angle is smaller in $La_{1.875}Ba_{0.125}CuO_4$ than in $La_{1.48}Nd_{0.4}Sr_{0.12}CuO_4$ [3]. At pressures greater than 5.0 GPa, where $La_{1.48}Nd_{0.4}Sr_{0.12}CuO_4$ is in the HTT phase, $T_c$ is seen to decrease nearly linearly with pressure at the rate $dT_c/dP = -0.78(8)$ K/GPa. For comparison, in optimally doped $La_{1.85}Sr_{0.15}CuO_4$ $T_c$ is found to decrease with pressure above 4.0 GPa at the somewhat higher rate $dT_c/dP = -1.2$ K/GPa [21], but the maximum $T_c$ is nearly twice as high for $La_{1.85}Sr_{0.15}CuO_4$ as for $La_{1.48}Nd_{0.4}Sr_{0.12}CuO_4$. It is plausible that the decrease of $T_c$ with pressure in the HTT phase is primarily due to the negative $dT_c/dP$ associated with *c*-axis compression [11, 14, 15]. The maximum value of $T_c(P)$ for $La_{1.48}Nd_{0.4}Sr_{0.12}CuO_4$ is approximately 22 K at 5 GPa for the HTT phase, whereas for a $La_{1.88}Sr_{0.12}CuO_4$ polycrystalline sample $T_c \approx 30$ K for the LTO1 phase at ambient pressure [2]. The lower maximum $T_c$ for $La_{1.48}Nd_{0.4}Sr_{0.12}CuO_4$ may be an effect of the increased disorder associated with Nd substitution, although a similar maximum value of $T_c$ has also been observed [11-13] for $La_{1.875}Ba_{0.125}CuO_4$ under hydrostatic pressure. Another possibility is the presence of residual charge order in $La_{1.48}Nd_{0.4}Sr_{0.12}CuO_4$ in the HTT phase (see discussion below).

The pressure dependence of $T_c$ to 1.0 GPa has recently been determined [14] *resistively* on a $La_{1.48}Nd_{0.4}Sr_{0.12}CuO_4$ crystal with the same nominal stoichiometry as our sample. The results of that study were interpreted to show that a hydrostatic pressure of 0.2 GPa is sufficient to stabilize the HTT phase in $La_{1.48}Nd_{0.4}Sr_{0.12}CuO_4$. Our structural and $T_c$ measurements, however, consistently find that a much higher pressure of 4.2 GPa is required to stabilize the HTT structure in this material. One plausible origin for this large discrepancy is the much greater sensitivity of resistivity measurements to the $T_c$ values in minority regions such as crystallographic twin boundaries, which may contain LTO2, LTO1 or HTT phases [23] with higher $T_c$ values than the bulk phase. A second possibility is that the pressure in that study was not fully hydrostatic, yielding an exceptionally rapid increase of $T_c$. A similar rapid rise of $T_c$ has been reported for LTT $La_{1.64}Eu_{0.2}Sr_{0.16}CuO_4$ under uniaxial pressure applied parallel to the (110) axis [15]: a pressure of only 0.3 GPa increased $T_c$ from $T_c$ = 10 K to ~17 K. A linear extrapolation of the data in reference [15] suggests that a uniaxial pressure of ~0.5 GPa along the (110) direction would increase $T_c$ to 22 K, the value we observe under a hydrostatic pressure of 5.0 GPa. If the uniaxial pressure suppresses the LTT phase in favor of the HTT phase, then the data in reference [15] suggest that a uniaxial pressure of only 0.5 GPa in the (110) direction is sufficient to accomplish this structural transformation! Structural measurements under uniaxial compression using single crystal x-ray or neutron diffraction are necessary to test this possibility.

The maximum $T_c$ of 22 K is found in the HTT phase of $La_{1.48}Nd_{0.4}Sr_{0.12}CuO_4$. This conclusion is consistent with results for $La_{1.88}Ba_{0.12}CuO_4$ [4, 11], as well as with other observations that suggest the highest superconducting transition temperatures occur when the Cu-O planes in lanthanum cuprates are flat, that is there are no tilt distortions [24]. Within the stripe scenario, the HTT structure has no pinning potential in the Cu-O planes to lead to charge ordering and consequent suppression of superconductivity. In addition, there is no Dzyaloshinsky-Moriya interaction in the HTT phase, and the interlayer magnetic exchange is fully frustrated. Thus the HTT structure should provide the most homogeneous and magnetically isotropic environment of the four tilt structures listed in Table I.

Very recent experimental work using scanning tunneling microscopy has highlighted the presence of charge ordering with (4*a* x 4*a*) or (4.5*a* x 4.5*a*) periodicities in the underdoped cuprates $Na_xCa_{2-x}CuO_2Cl_2$ [6] and $Bi_2Sr_2CaCu_2O_{8+\delta}$ [7], respectively. Several theoretical models [8-10] have been proposed to explain these new observations. Each of these models invokes some type of crystallization of the holes [10], or of singlet pairs of holes [8, 9], in the Cu-O plane. If the charges condense as singlet pairs, it is possible that such a charge-ordered structure also exists in the LTT phase of $La_{1.48}Nd_{0.4}Sr_{0.12}CuO_4$ at x = 1/8, leading to the suppression of superconductivity at ambient pressure [8]. Our data show that application of hydrostatic pressure reduces the LTT distortion and increases the superconducting transition temperature. Of course, it is interesting to ask what is the effect of pressure on the charge ordering itself?

Measurement of $T_c$ vs. x in a series of samples of composition $La_{1.6-x}Nd_{0.4}Sr_xCuO_4$, under sufficient pressure to stabilize the HTT phase, would show whether the superconducting $T_c = 22$ K at x = 1/8 is anomalously low compared with samples with smaller or large x values. If so, this would provide indirect evidence for residual charge order in the HTT phase of $La_{1.48}Nd_{0.4}Sr_{0.12}CuO_4$. This type of data has been presented [13] for $La_{2-x}Ba_xCuO_4$, and does show a local minimum of $T_c$ at x = 1/8. Unfortunately, high-pressure structural data were not presented in that work, nor were the superconducting $T_c$s measured to pressures high enough to observe a clear maximum value of $T_c$ (as we see for $La_{1.48}Nd_{0.4}Sr_{0.12}CuO_4$ in Figure 6). Thus we believe it is not safe to conclude at this time that there is a local minimum of $T_c$ in the HTT phase of $La_{2-x}Ba_xCuO_4$, or any other lanthanum cuprates, when x = 1/8. Clarifying this issue is, however, an important goal for future high-pressure experiments. In addition, it would be very interesting to directly observe, by neutron or x-ray scattering, the charge-ordered state in these materials as a function of pressure. Such measurements will be difficult, however, because the charge-order superlattice peaks are weak in the LTT phase [5, 25], and must be detected in the presence of the background scattering from the diamond anvil or other type of pressure cell. It is also well-known that $La_{2-x}Sr_xCuO_4$ at ambient pressure shows a local minimum of $T_c$ at x = 1/8 [26-28], suggesting that some charge and spin order exists in the LTO1 phase of this material as well. In fact, incommensurate elastic magnetic peaks have been reported [29] in neutron-scattering measurements for $La_{1.88}Sr_{0.12}CuO_4$, although no direct evidence yet exists for charge order. In total, these results reinforce the unique nature of the electronic state at x = 1/8 in the lanthanum cuprates, and the strong coupling of this state to the structural distortions, although we still lack a complete understanding of these phenomena.

Neutron diffraction studies [30] of $La_{2-x}Sr_xCuO_4$ and $La_{1.6-x}Nd_{0.4}Sr_xCuO_4$ single crystals have clearly shown that the charge and spin order is directly influenced by both the tilt magnitude and tilt direction of the $CuO_6$ octahedra. The effect of pressure on the superconducting $T_c$ of $La_{1.48}Nd_{0.4}Sr_{0.12}CuO_4$ must ultimately be understood in terms of the detailed relationships between structure, charge and spin order, and superconductivity. Toward this end, our x-ray diffraction and AC susceptibility measurements provide the first definitive picture of the structural and superconducting properties of $La_{1.48}Nd_{0.4}Sr_{0.12}CuO_4$ under nearly ideal hydrostatic pressure conditions.

*Acknowledgements* We thank Robert J. Smalley (DuPont) for help with diamond anvil cell pressure calibration and Asa Hopkins (NHMFL) for loading samples in the pressure cells used for the x-ray diffraction measurements. Work at Washington University was supported by NSF grant DMR-0101809.

4. REFERENCES

1. J.D Axe, A.H. Moudden, D. Hohlwein, D.E. Cox, K.M. Mohanty, A.R. Moodenbaugh, and Y. Xu, Phys. Rev. Lett. **62**, 2751 (1989).

2. M.K. Crawford, R.L. Harlow, E.M. McCarron, W.E. Farneth, J.D. Axe, H.E. Chou, and Q. Huang, Phys. Rev. B **44**, 7749 (1991).

3. J.D. Axe and M.K. Crawford, J. Low Temp. Phys. **95**, 271 (1994).

4. M.K. Crawford, R.L. Harlow, E.M. McCarron, S.W. Tozer, Q. Huang, D.E. Cox and Q. Zhu, in *High-$T_c$ Superconductivity 1996: Ten Years after the Discovery*, ed. E Kaldis et al., p. 281 (1997).

5. J.M. Tranquada, B.J. Sternlieb, J.D. Axe, Y. Nakamura, and S. Uchida, Nature **375**, 561 (1995); N. Ichikawa, S. Uchida, J.M. Tranquada, T. Niemöller, P.M. Gehring, S.-H. Lee, and J.R. Schneider, Phys. Rev. Lett. **85**, 1738 (2000).

6. T. Hanaguri, C. Lupien, Y. Kohsaka, D.-H. Lee, M. Azuma, M. Takano, H. Takagi, and J.C. Davis, Nature **430**, 1001 (2004).

7. K. McElroy, D.-H. Lee, J.E. Hoffman, K.M. Lang, E.W. Hudson, H. Eisaki, S. Uchida, J. Lee, and J.C. Davis, cond-mat/0404005.

8. P.W. Anderson, cond-mat/0406038.

9. H.-D. Chen, O. Vafek, A. Yazdani, and S.C. Zhang, cond-mat/0402323.

10. H.C. Fu, J.C. Davis, and D.-H. Lee, cond-mat/0403001.

11. H. Takahashi and N. Mori in: *Studies of High Temperature Superconductors*, Vol. 16/17, ed. A.V. Narlikar (Nova Science Publishers, Inc., N.Y., 1995) p. 1; J.S. Schilling and S. Klotz, in *Physical Properties of High Temperature Superconductors,* edited by D.M. Ginsberg (World Scientific, Singapore, 1992), Vol. III, p. 59.

12. S. Katano, S. Funahashi, N. Môri, Y. Ueda, and J.A. Fernandez-Baca, Phys. Rev. B **48**, 6569 (1993).

13. N. Yamada and M. Ido, Physica C **203**, 240 (1992).

14. S. Arumugam, N. Mori, N. Takeshita, H. Takashima, T. Noda, H. Eisaki, and S. Uchida, Phys. Rev. Lett. **88**, 247001 (2002).

15. N. Takeshita, T. Sasagawa, T. Sugioka, Y. Tokura, and H. Takagi, J. Phys. Soc. Jpn. **73**, 1123 (2004).

16. B. Simovič, M. Nicklas, P.C. Hammel, M. Hücker, B. Büchner, and J.D. Thompson, Europhys. Lett. **66**, 722 (2004).

17. A.H. Morrish, *Physical Principles of Magnetism*, John Wiley and Sons, New York (1965).

18. J. Fischer, V. Radeka, and G.C. Smith, IEEE Trans. Nucl. Sci. **NS-33**, 136 (1986); G.C. Smith, Sync. Rad. News **4**, 24 (1991

19. The $T_c$ for Pb is known to decrease linearly under pressure to 5 GPa at the rate $dT_c/dP \cong$ -0.365 K/GPa (see A. Eiling and J.S. Schilling, J. Phys. F **11**, 623 (1981)); at higher pressures a $T_c(P)$ calibration for Pb in $^4$He pressure medium was used (J. Thomasson, C. Ayache, I.L. Spain, M. Villedieu, J. Appl. Phys. **68**, 5933 (1990)).

20. J.S. Schilling, Mat. Res. Soc. Symp. Proc. **22**, 79 (1984); J.S. Schilling, J. Diederichs, S. Klotz, and R. Sieburger, in *Magnetic Susceptibility of Superconductors and Other Spin Systems,* edited by R.A. Hein, T.L. Francavilla, and D.H. Liebenberg (Plenum, New York, 1991), p. 107.

21. A.C. Larson and R.B. Von Dreele, GSAS-General Structure Analysis System Report LA-UR-86-748 (1987). Los Alamos National Laboratory.

22. N. Mori, C. Murayama, H. Takahashi, H. Kaneko, K. Kawabata, Y. Iye, S. Uchida, H. Takagi, Y. Tokura, Y. Kubo, H. Sasakura, K. Yamaya, Physica C **185-189**, 40 (1991).

23. Y. Zhu, A.R. Moodenbaugh, Z.X. Cai, J. Tafto, M. Suenaga, and D.O. Welch, Phys. Rev. Lett. **73**, 3026 (1994).

24. H. Takahashi, H. Shaked, B.A. Hunter, P.G. Radaelli, R.L. Hitterman, D.G. Hinks, and J.D. Jorgensen, Phys. Rev. B **50**, 3221 (1994); B. Dabrowski, Z. Wang, K. Rogacki, J.D. Jorgensen, R.L. Hitterman, J.L. Wagner, B.A. Hunter, P.G. Radaelli, and D.G. Hinks, Phys. Rev. Lett. **76**, 1348 (1996).

25. T. Niemöller, N. Ichikawa, T. Frello, H. Hünnefeld, N.H. Andersen, S. Uchida, J.R. Schneider, and J.M. Tranquada, Eur. Phys. J. B-Condensed Matter **12**, 509 (1999); T. Niemöller, H. Hünnefeld, T. Frello, N.H. Andersen, N. Ichikawa, S. Uchida, and J.R. Schneider, J. Low Temp. Phys. **117**, 455 (1999).

26. H. Takagi, T. Ido, S. Ishibashi, M. Uota, S. Uchida, and Y. Tokura, Phys. Rev. B **40**, 2254 (1989).

27. M.K. Crawford, W.E. Farneth, E.M. McCarron, R.L. Harlow, and A.H. Moudden, Science **250**, 1390 (1990).

28. A.R. Moodenbaugh, L.H. Lewis, and S. Soman, Physica C 290, **98** (1997).

29. H. Kimura, K. Hirota, H. Matsushita, K. Yamada, Y. Endoh, S.-H. Lee, C.F. Majkrzak, R. Erwin, G. Shirane, M. Greven, Y.S. Lee, M.A. Kastner, and R.J. Birgeneau, Phys. Rev. B **59**, 6517 (1999).

30. See, for example, S. Wakimoto, J.M. Tranquada, T. Ono, K.M. Kojima, S. Uchida, S.-H. Lee, P.M. Gehring, R.J. Birgeneau, Phys. Rev. B. **64**, 174505 (2001); S.

Wakimoto, R.J. Birgeneau, M.A. Kastner, Y.S. Lee, R. Erwin, P.M. Gehring, S.-H. Lee, M. Fujita, K. Yamada, Y. Endoh, K. Hirota, G. Shirane, Phys. Rev. B **61**, 3699 (2000).

**Table I.** Descriptions of the four tilt structures observed in $La_{1.48}Nd_{0.4}Sr_{0.12}CuO_4$. The order parameters $Q_1$ and $Q_2$ are defined in Figure 1.

| Structure[a] | Space Group | Order Parameters |
|---|---|---|
| HTT | *I4/mmm* | $Q_1 = Q_2 = 0$ |
| LTO1 | *Bmab*[b] | $|Q_1| \neq 0$ or $|Q_2| \neq 0$ |
| LTO2 | *Pccn* | $|Q_1| \neq |Q_2| \neq 0$ |
| LTT | *P4$_2$/ncm* | $|Q_1| = |Q_2| \neq 0$ |

[a] HTT: high temperature tetragonal; LTO1: low temperature orthorhombic 1; LTO2: low temperature orthorhombic 2; LTT: low temperature tetragonal

[b] This is a nonstandard setting for the space group *Cmca*, chosen to maintain the *c*-axis as the long axis of the unit cell.

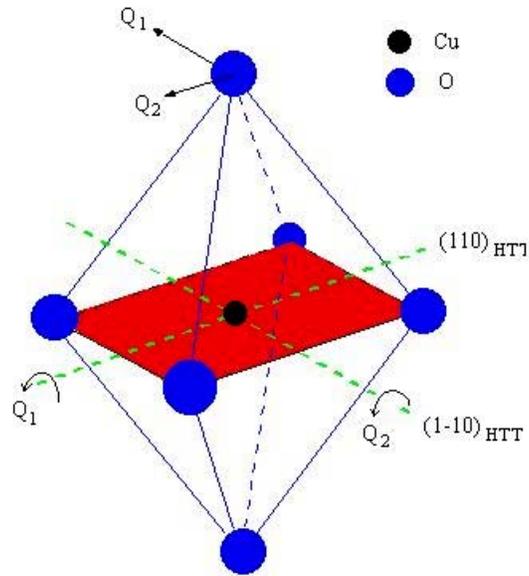

Figure 1. A schematic view of a $CuO_6$ octahedron illustrating the relative orientations of the (110) and (1-10) axes of the HTT structure. The LTO1 structure is obtained by a single rotation of magnitude $|Q_1|$ or $|Q_2|$ about one of these axes (yielding one of the two orthorhombic twin structures). The LTO2 structure is obtained by simultaneous unequal rotations of magnitude $|Q_1|$ and $|Q_2|$ about both axes (where $|Q_1| \neq |Q_2|$). The LTT structure has $|Q_1| = |Q_2| \neq 0$.

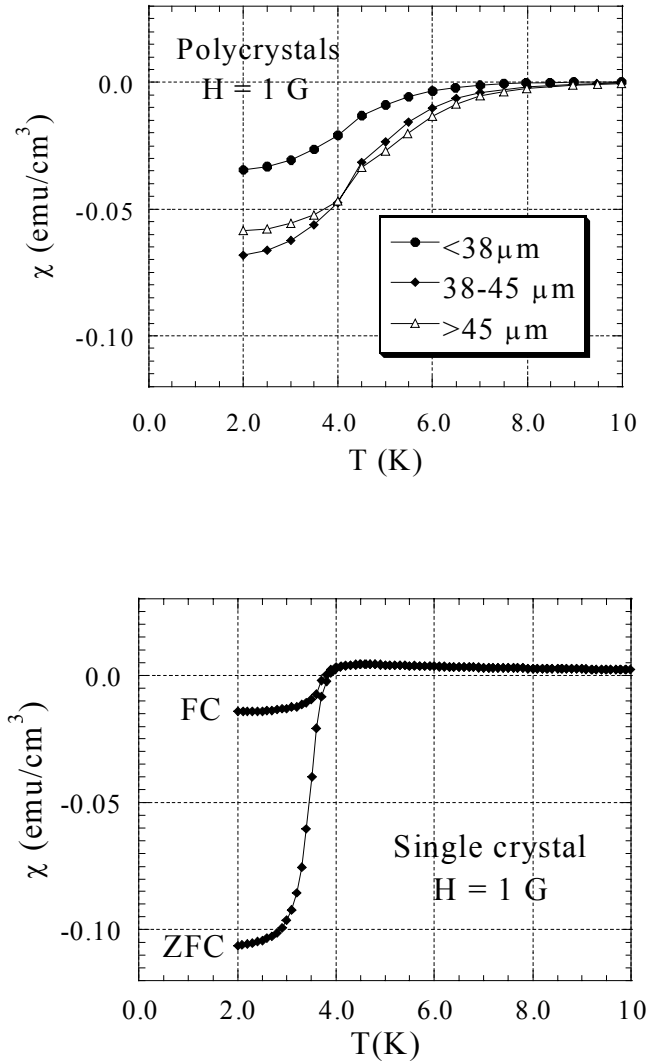

Figure 2. (top) DC volume magnetization versus temperature at 1 G applied magnetic field for the $La_{1.48}Nd_{0.4}Sr_{0.12}CuO_4$ polycrystalline sample studied in the present high-pressure x-ray powder diffraction experiments. Only zero-field-cooled (ZFC) data is shown. The three curves represent data from different particle size distributions obtained after passing the sample through various metal mesh sieves. (bottom) DC volume magnetization versus temperature at 1 G applied magnetic field for the $La_{1.48}Nd_{0.4}Sr_{0.12}CuO_4$ single crystal studied in the present high-pressure magnetic

susceptibility experiments. Zero-field-cooled (ZFC) and field-cooled (FC) data are shown. The data were not corrected for demagnetization effects. The crystal orientation with respect to the applied magnetic field was unknown.

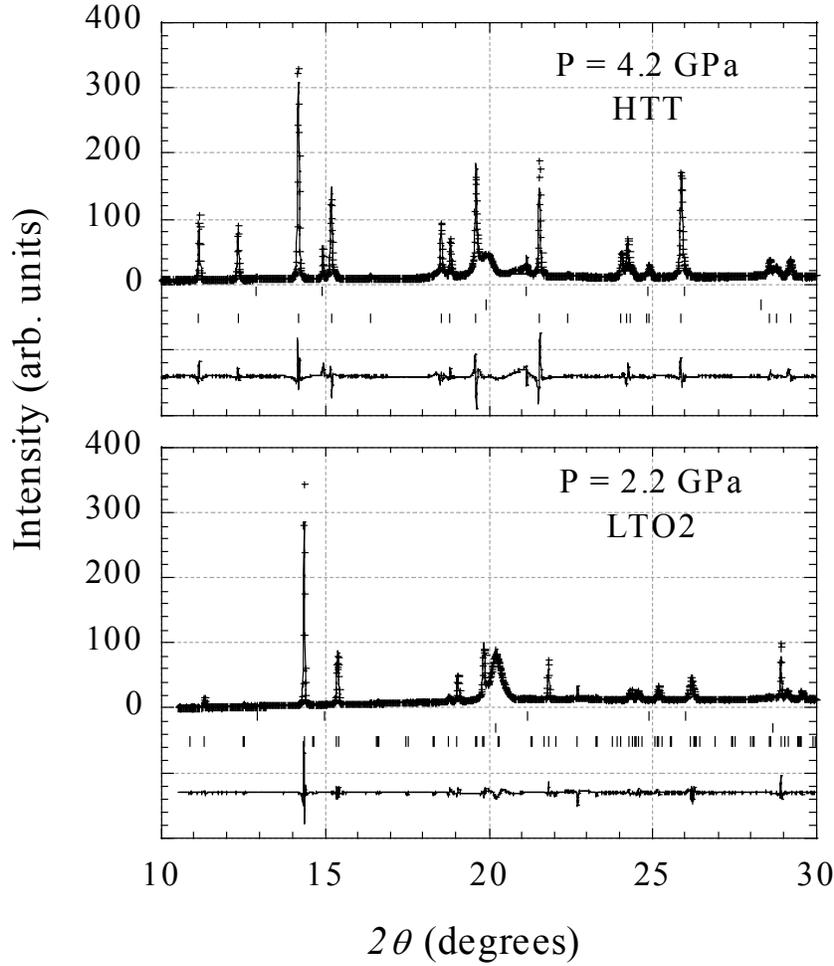

Figure 3. (top) X-ray powder diffraction scan and GSAS refinement of $La_{1.48}Nd_{0.4}Sr_{0.12}CuO_4$ at a pressure of 4.2 GPa and a temperature of 10 K. The x-ray wavelength was 0.7000(9) Å. In the upper trace the crosses are the data, and the solid line is the calculated pattern; the lowest trace is the residual. The vertical tick marks are located at the positions of (top) NaCl included in the diamond anvil cell as an internal manometer, (middle) Fe due to the diamond anvil cell gasket, and (bottom) $La_{1.48}Nd_{0.4}Sr_{0.12}CuO_4$. The HTT unit cell parameters at T = 10 K are $a = b = 3.7429(2)$ Å

and $c = 13.0285(9)$ Å. (bottom) X-ray powder diffraction scan and GSAS refinement of $La_{1.48}Nd_{0.4}Sr_{0.12}CuO_4$ at a pressure of 2.2 GPa and a temperature of 10 K. The x-ray wavelength was 0.7080(5) Å. The symbols are the same as in the top frame. The LTO2 unit cell parameters at T = 10 K are $a = 5.2934(3)$ Å, $b = 5.3100(3)$ Å, and $c = 13.0477(8)$ Å.

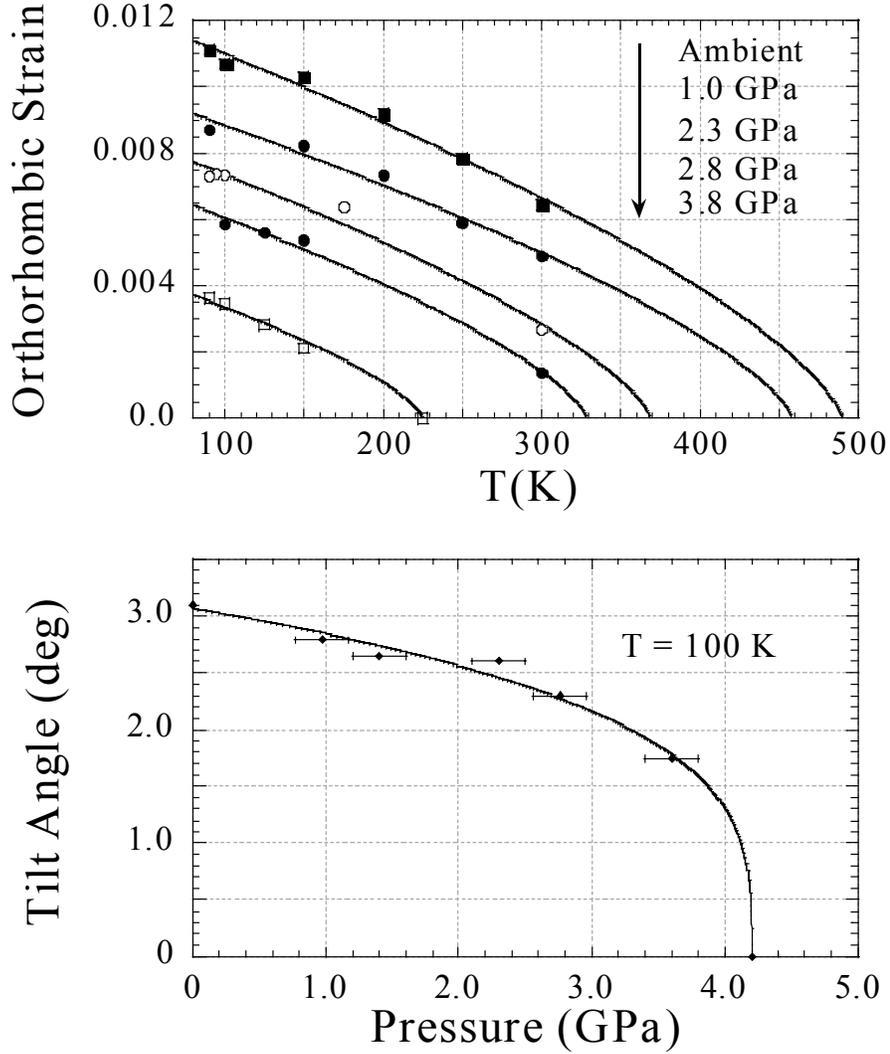

Figure 4.   (top) Orthorhombic strain $[(b-a)/a]$ as a function of pressure for $La_{1.48}Nd_{0.4}Sr_{0.12}CuO_4$.  The solid lines are fits to $[(b-a)/a] \propto (T_0-T)^{2\beta}$, where $T_0$ is the HTT $\rightarrow$ LTO1 structural phase transition temperature and the exponent $2\beta = 0.7$ is fixed to the value characteristic of the 3D, $n = 2$, XY model.  (bottom) $CuO_6$ octahedra tilt angle as a function of pressure at a temperature of 100 K.  The solid line is a fit to the expression $Q_1(P) \propto (P_0 - P)^{\alpha}$, where $Q_1(P)$ is the tilt angle, $\alpha = 0.28$, and $P_0 = 4.2$ GPa.

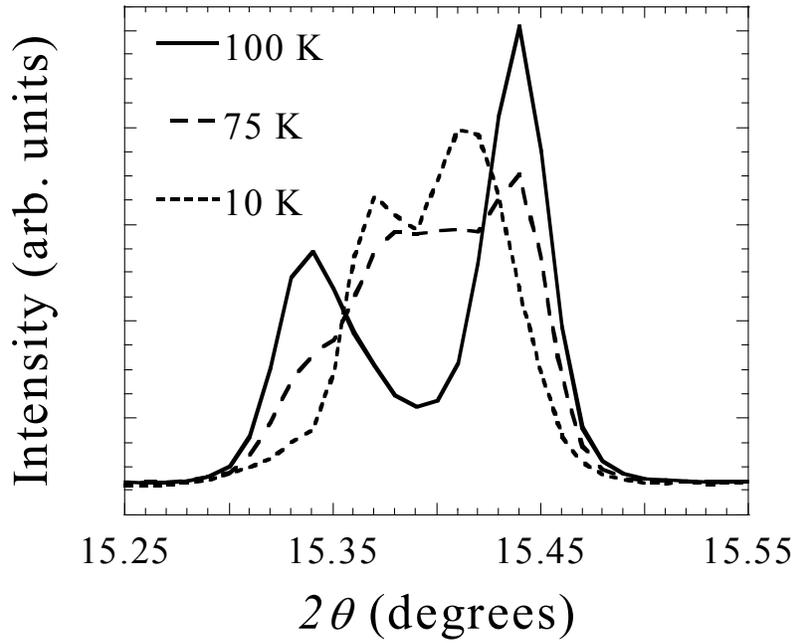

Figure 5. The $(110)_{HTT}$ Bragg reflection of $La_{1.48}Nd_{0.4}Sr_{0.12}CuO_4$ at three temperatures at a pressure of 2.2 GPa. The x-ray wavelength was 0.7080(5) Å. At T = 100 K the structure is LTO1, at 75 K LTO1 + LTO2, and at 10 K LTO2.

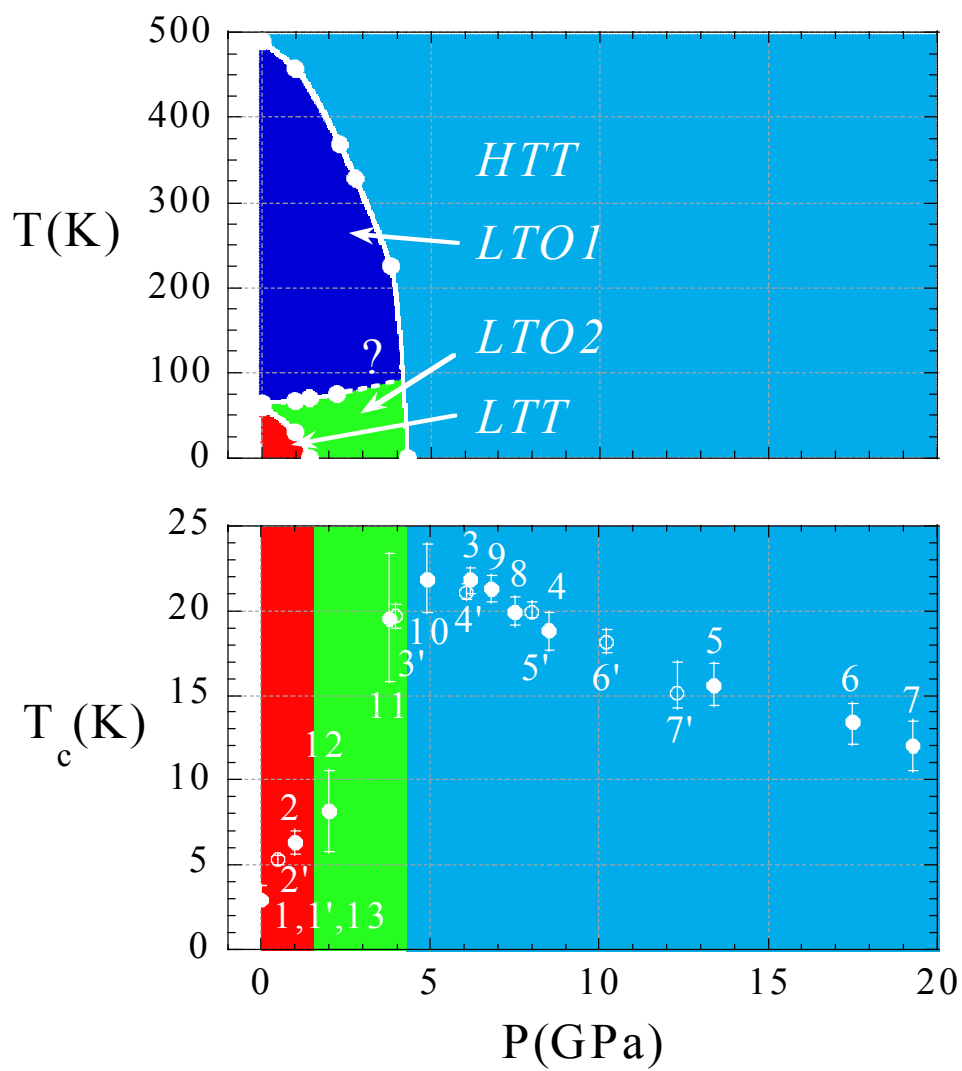

Figure 6. (top) Pressure-temperature phase diagram for $La_{1.48}Nd_{0.4}Sr_{0.12}CuO_4$ determined by high-pressure x-ray diffraction. The LTT phase region is shown in red, the LTO2 region in green, the LTO1 region in dark blue, and the HTT region in light blue. Note the increase (~ 10 K from 1 bar to 2.2 GPa) in the first-order LTO1 → LTO2 transition temperature with pressure. Above 2.2 GPa it becomes difficult to detect the presence of an LTO1 → LTO2 transition because of the increasingly small orthorhombic strain of the LTO1 structure; this region is indicated in the figure by a question mark. (bottom)

Superconducting transition temperature ($T_c$) versus pressure (P) for $La_{1.48}Nd_{0.4}Sr_{0.12}CuO_4$. The color coding shows the crystal structure, as in the top frame. The numbers give the order of measurement for the two different high-pressure experiments (primed open circles (o) and unprimed filled circles (●)). The vertical error bars give the transition widths.